\documentclass[aps,prb,twocolumn,superscriptaddress,showpacs]{revtex4}
\usepackage{amssymb}
\usepackage{graphicx}
\usepackage{natbib}
\usepackage[usenames]{color}
\usepackage[dvipsnames]{xcolor}

\def\C60{C$_{60}$}
\def\Mg5C60{Mg$_5$C$_{60}$}
\def\A4C60{A$_4$C$_{60}$}
\def\cm-1{cm$^{-1}$}
\def\T1u{$T_{1u}$}

\begin{document}

\title{Mg$_5$C$_{60}$: A stable two dimensional conducting polymer }
\author{D. Quintavalle}
\thanks{Corresponding author. Electronic address: dario@esr.phy.bme.hu}
\affiliation{Budapest University of Technology and Economics, Institute of Physics and Condensed Matter Physics Research Group of the Hungarian Academy of Sciences, H-1521, Budapest P.O.Box 91, Hungary}
\author{F. Borondics}
\altaffiliation{Present address: Advanced Light Source Division, Lawrence Berkeley National
Laboratory, Berkeley, CA 94720-8226}
\affiliation{Research Institute for Solid State Physics and Optics, Hungarian Academy of
Sciences, P.O. Box 49, Budapest, Hungary H-1525}
\author{G. Klupp}
\affiliation{Research Institute for Solid State Physics and Optics, Hungarian Academy of
Sciences, P.O. Box 49, Budapest, Hungary H-1525}
\author{A. Baserga}
\affiliation{NEMAS-Center for NanoEngineered MAterials and Surfaces, Dipartimento di
Ingegneria Nucleare, Politecnico di Milano, I-20133 Milano,
Italy}
\author{F. Simon}
\affiliation{Budapest University of Technology and Economics, Institute of Physics and Condensed Matter Physics Research Group of the Hungarian Academy of Sciences, H-1521, Budapest P.O.Box 91, Hungary}
\author{A. J\'{a}nossy}
\affiliation{Budapest University of Technology and Economics, Institute of Physics and Condensed Matter Physics Research Group of the Hungarian Academy of Sciences, H-1521, Budapest P.O.Box 91, Hungary}
\author{K. Kamar\'{a}s}
\affiliation{Research Institute for Solid State Physics and Optics, Hungarian Academy of
Sciences, P.O. Box 49, Budapest, Hungary H-1525}
\author{S. Pekker}
\affiliation{Research Institute for Solid State Physics and Optics, Hungarian Academy of
Sciences, P.O. Box 49, Budapest, Hungary H-1525}

\begin{abstract}
We present a study on the structural, spectroscopic, conducting and magnetic properties of
Mg$_{5}$C$_{60}$, a two dimensional (2D) fulleride polymer. The polymer
phase is stable up to the exceptionally high temperature of 823 K. Infrared and Raman studies suggest the
formation of single bonds between fulleride ions and possibly Mg - C$_{60}$
covalent bonds. 
Mg$_{5}$C$_{60}$ is a metal at ambient temperature as shown by electron spin
resonance and microwave conductivity measurements. The smooth transition
from a metallic to a paramagnetic insulator state below 200~K is attributed to
Anderson localization driven by structural disorder.
\end{abstract}

\maketitle

%\begin{keyword}

%\end{keyword}

% main text

\section{Introduction}

The unusual physical and structural properties of alkali metal intercalated
fulleride polymers have received considerable attention since the discovery in 1994
of the linear AC$_{60}$ (A=K, Rb, Cs) conducting polymers. \cite{pekker94}
Chemical reactions between charged fulleride ions in solids are rather
common. The polymer phases usually form spontaneously or under mild pressure
from the monomeric crystalline salts. Polymers of C$_{60}^{n-}$ anions are
usually unstable above temperatures of about 400 K. The depolymerization is reversible, in contrast to
photopolymerization \cite{Rao93} and pressure polymerization of neutral C$%
_{60}$ (Refs. \onlinecite{Iwasa94,Nunez95}). According to quantum chemical
calculations, \cite{Pekker99} [2+2] cycloaddition is favored in A$_{n}$C$%
_{60}$ for low values of $n$, and configurations with single interfullerene bonds are more stable
for $n\geq 3$. In agreement with these expectations,
the stable form of AC$_{60}^{-}$ is a cycloaddition polymer,\cite{steph94}
while single intermolecular bonds occur between C$_{60}^{n-}$ anions in
Na$_{4} $C$_{60}$ (Ref. \onlinecite{oszi97}) and Na$_{2}$RbC$_{60}$ (Ref. %
\onlinecite{Bendele98b}). Single and double bonds can appear simultaneously,
as has been recently demonstrated in Li$_{4}$C$_{60} $ (Refs. %
 \onlinecite{Margadonna04, Ricco05}) where single bonds connect polyfulleride chains
held together by [2+2] cycloaddition bonds.

Fulleride polymers have small electronic bandwidths and large on-site
electron repulsion and thus are strongly correlated electron systems. Small
differences in the lattice parameters and/or variations in the chain
orientation \cite{Launois98} can change the ground state profoundly. For
example, the linear polymers KC$_{60}$ and Na$_{2}$RbC$_{60}$ have metallic
ground states while RbC$_{60}$ and CsC$_{60}$ with the same type of polymer
chains but different chain orientations undergo a metal-insulator transition
to an antiferromagnetic spin density wave ground state.\cite{bommeli95}
Electron-electron correlations play an important role in 2D fulleride
polymers as well: Na$_{4}$C$_{60}$ is a strongly correlated metal \cite%
{oszi97} while Li$_{4}$C$_{60}$ (Ref. \onlinecite{Ricco05}) is a nonmagnetic
insulator.

In this paper, we present a study of the structure and physical properties
of the recently synthesized fulleride polymer Mg$_{5}$C$_{60}$. In a
previous study of Mg$_{x}$C$_{60}$, a stoichiometric compound \cite%
{Borondics03} was reported for $x=4$. In the present work we improved the
synthesis and conclude from a series of samples with varying Mg content that
the homogeneous phase lies in a range of Mg concentrations between $x=5$
and $x=5.5$. Mg$_{5}$C$_{60}$ is the only example
of an alkaline earth fulleride polymer. Previous studies of alkaline earth
fullerides focused mostly on Ca$_{x}$C$_{60}$ and Ba$_{x}$C$_{60}$ (Refs. %
\onlinecite{Kortan92,Baenitz95}) superconductors which are not polymers. An
early study of Mg doped fulleride films reported an insulating behavior for
all Mg concentrations \cite{Chen92} in disagreement with our present results.

We find that Mg$_{5}$C$_{60}$ is a 2D polymer which is
metallic at high temperature and undergoes a gradual transition to an
insulating ground state as the temperature is lowered. It is stable up to the remarkably high temperature of 823 K.

\section{Experimental}

\begin{figure}[tbp]
\centerline{\includegraphics[width=1\hsize,trim=0 0 12pt 0]{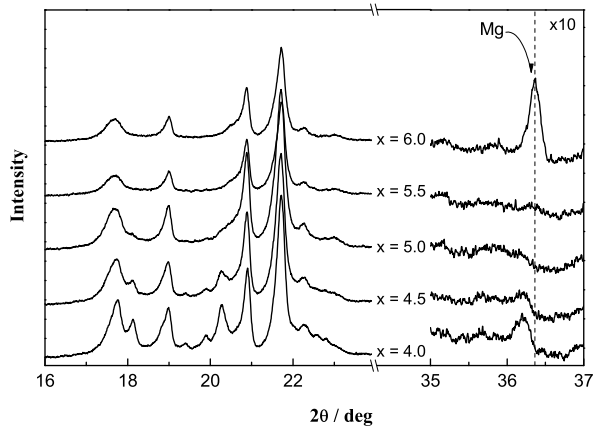}}
\caption{ X-ray powder diffraction data for a series of Mg$_x$C$_{60}$ polymers, with nominal concentrations $x$ varying from 4.0 to 6.0. Dotted line above 36$^{\circ }$ corresponds to Mg metal. }
\label{fig:XRAY4}
\end{figure}

Samples of Mg$_{x}$C$_{60}$ with nominal concentrations $x$ = 4 to 6, in
steps of $\Delta x=0.5$, were prepared by solid state reaction between C$%
_{60}$ and pure Mg powders under argon atmosphere in a dry box. The powder mixture
was placed in carbon steel containers to avoid
reaction between Mg grains and quartz vessels. Mg grain surface activation
at 753 K was followed by several annealing steps at temperatures from 653~K
to 723~K. To improve the homogeneity of the sample, we used Mg powder with
smaller grain sizes than in the previous preparation\cite{Borondics03}.
Powders were reground before each annealing step.

X-ray powder diffraction was performed using a HUBER G670
Guinier image plate camera in transmission mode and highly monochromatic CuK$%
\alpha _{1}$ radiation. Infrared spectra were recorded on pressed KBr
pellets with a Bruker IFS 28 FTIR spectrometer under dynamic vacuum. Raman
spectra were collected using a T64000 Jobin-Yvon spectrometer in triple
grating configuration with spectral resolution better than 3~cm$^{-1}$. The
532 nm line of a frequency-doubled Nd-YAG laser was the excitation source.
The diameter of the spot was 100~$\mu $m and the nominal
irradiance about 150~W/cm$^{2}$, corresponding to $\sim$ 0.1 mW power on the sample.

Microwave conductivity was measured at 10~GHz on fine Mg$_{5}$C$_{60}$
powders mixed with high purity SnO$_{2}$ powder to electrically isolate the
grains. A 10 GHz copper cylindrical TE${011}$ resonant cavity was used for
the cavity perturbation technique conductance measurements \cite%
{nebendahl01,buravov71}. In this method changes in the quality factor are
measured using lock-in detection of the frequency modulated response. The
sample is placed at the cavity center where the microwave magnetic field is
maximum and microwave currents encircle the grains. This geometry is well
adapted for fine powders as the microwave electric field is not shielded by
depolarization effects. \cite{buravov71} The typical grain size was much
less than the microwave penetration depth. The temperature dependence of the
conductivity $\sigma (T)$, is proportional to $1/Q-1/Q_{0}$.  $Q$ and $Q_{0}$
, the quality factors of the cavity with and without the sample, respectively, were measured
in two separate runs. $Q_{0}$ was about $16000$ and it changed little with
temperature. Ohmic losses of a single spherical grain with radius $r$ and
conductivity $\sigma $ are proportional to $r^{5}\sigma $. The determination
of the absolute value of the conductivity is not possible without a detailed
knowledge of the distribution of $r$.

\begin{figure}[tbp]
\centerline{\includegraphics[width=1\hsize,trim=0 0 12pt 0]{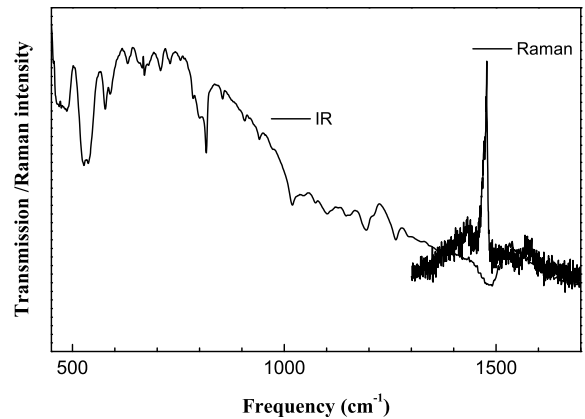}}
\caption{ Infrared and Raman spectra of Mg$_{5}$C$_{60}$.}
\label{fig:IRRamanb}
\end{figure}

Electron spin resonance (ESR)
spectroscopy was performed at 9 and 225~GHz on samples sealed in quartz
tubes under 200~mbar He pressure. At 9~GHz a commercial Bruker ELEXSYS 500
spectrometer was used, the 225~GHz spectra were recorded by a home-built
spectrometer. The spin susceptibility was obtained by integrating the ESR
spectra. MgO doped with 1.5~ppm of Mn$^{2+}$ ions was the ESR intensity reference to
obtain the susceptibility between 100 and 300~K. Below 100~K the Mn:MgO
ESR saturates and fine orthorhombic polymeric KC$_{60}$ powder (a low
conductivity metal with a temperature independent susceptibility) was used
as an intensity reference.

\section{Results and discussion}

\subsection{X-ray diffraction}

Powder diffractograms of a series of Mg$_{x}$C$_{60}$ salts (Fig.~\ref%
{fig:XRAY4}) show that the samples are single-phase for nominal
concentration values.  The peak above 36$^{\circ }$ demonstrates
that above $x=5.5$ nominal composition some Mg metal is present. The single-phase material has
been indexed in the previous paper\cite{Borondics03} as rombohedral with
lattice parameters $a=b=9.22 \mathring{A},  c=25.25 \mathring{A}, \gamma
=120^{\circ }$. These values indicate the formation of two-dimensional
hexagonal C$_{60}$ polymeric sheets in the $ab$ plane. The material belongs
to the $R\overline{3}m$ space group. C$_{60}$ molecules are positioned
at the fractional coordinates (0,0,0), (2/3,1/3,1/3) and (1/3,2/3,2/3),
while the Mg positions are (0,0,0.23) and (0,0,0.43). This structure
corresponds to the nominal composition $x=4$. We regard the magnesium content presented in Fig.~\ref{fig:XRAY4} more reliable than the previous one \cite{Borondics03}, due to the improved synthesis method using fine-powered magnesium metal. The x-ray results also indicate that Mg$_5$C$_{60}$ is not a line phase but the structure presented here and in Ref.~\onlinecite{Borondics03} comprises a range of stoichiometries between x=5 and 5.5.  From the structural data,
however, we cannot determine the positions of the additional magnesium ions.
A range of
alkali concentrations without changes in the structure of the
polymer has been observed in Li$_{x}$C$_{60}$ ($3\leq x\leq 5$),\cite{Ricco05}
with the structural parameters given by a Rietveld fit corresponding to $x=4$.
Subsequent photoemission data on Li$_{x}$C$_{60}$ thin films with the same structure proved the
charge transfer to be incomplete and constant throughout the homogeneity range.\cite{macovez07}

\subsection{Infrared and Raman spectroscopy}

\begin{figure}[tbp]
\centerline{\includegraphics[width=1\hsize,trim=0 0 12pt 0]{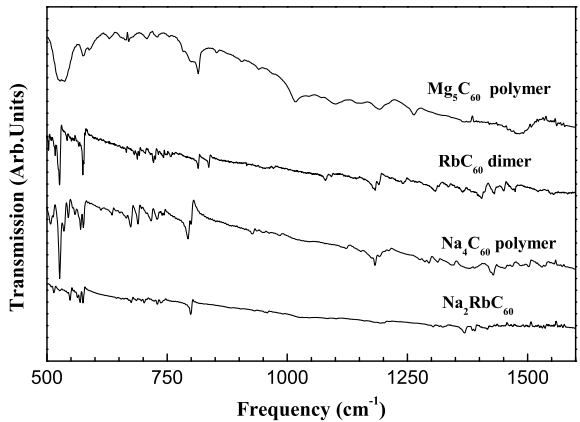}}
\caption{ Comparison of the infrared spectra of various singly bonded
fulleride oligo- and polymers.}
\label{fig:PolymersIR}
\end{figure}

The infrared and Raman spectra of Mg$_{5}$C$_{60}$ are shown in Fig.~\ref%
{fig:IRRamanb}. The low-frequency part corresponds to a symmetry-reduced C$%
_{60}$ molecule with the principal $T_{1u}$ modes at 523, 576 and 1193~cm,
and several new modes, especially in the region between 700-900~cm$^{-1}$.
The lower symmetry confirms polymerization in accordance with x-ray
diffraction. The line around 815~cm$^{-1}$ is of particular interest. This
line appears in the spectrum of all C$_{60}$-based polymers\cite%
{tagma2000,klupp03} where the C$_{60}$ molecules are connected by single
bonds (see~Fig.~\ref{fig:PolymersIR}). We consider the appearance of this
characteristic peak in Mg$_{5}$C$_{60}$ as proof for single bonds between
fulleride ions.

The high-frequency region around the $T_{1u}$(4) mode shows an unusual pattern, dominated
by a broad feature around 1480~cm$^{-1}$. This feature cannot be explained
by a shift in the $T_{1u}$(4) mode by charge transfer from the Mg atom,
neither by polymerization effects, which are both known to reduce the
vibrational frequencies. \cite{rao97,wagberg02} Instead, we assign this peak to the A$_g$(2) vibrational mode, which is Raman active in undistorted C$_{60}$ . Our assignment is supported by the fact that the peak is also dominant in the Raman spectrum. A similar effect was reported for C$_{60}$ monolayers adsorbed on metal \cite{rudolf02} and semiconductor surfaces. \cite{dumas96} In those systems, an intense infrared absorption above the
highest-frequency $T_{1u}$\ mode is observed, together with a Raman counterpart, \cite{peremans97} attributed to the $A_{g}$(2)
mode rendered infrared active by symmetry reduction and
amplified by the "vibrational phase relaxation" mechanism.\cite{erley89}
Prerequisites for this mechanism are 1. a free-electron continuum, 2.
low-frequency vibrational or translational modes which mediate the coupling.
According to our ESR and microwave data (see below), conducting electrons
are present in Mg$_{5}$C$_{60}$ and it is known from Raman spectroscopy of C$%
_{60}$ photopolymers that intermolecular vibrations appear around 100~cm$%
^{-1}$ (Ref. \onlinecite{Rao93}), thus both criteria are met. In the case of
Mg$_{5}$C$_{60}$ shown in Fig.~\ref{fig:IRRamanb} the mode is both infrared
and Raman-active, but, unlike References \onlinecite{rudolf02} and \onlinecite{dumas96}, in both cases shifted up by 12~cm$^{-1}$ compared to
the 1468~cm$^{-1}$ pristine $A_{g}$(2) mode. Such blue shift, albeit
smaller, has been observed in C$_{60}$ monolayers adsorbed on platinum, \cite%
{cepek96} where it was explained by covalent bonding to the metal, resulting
in electron back transfer from the fullerene to the metal surface. Covalent bonding to at least some of the Mg ions cannot be excluded in the present case either, considering the close values of the Mg ionization potential and the LUMO level of C$_{60}$. The phenomenon described here warrants further investigation since it was not seen in bulk fulleride salts before.

\subsection{Microwave conductivity and ESR}

\begin{figure}[tbp]
\centerline{\includegraphics[width=1\hsize,trim=0 0 12pt 0]{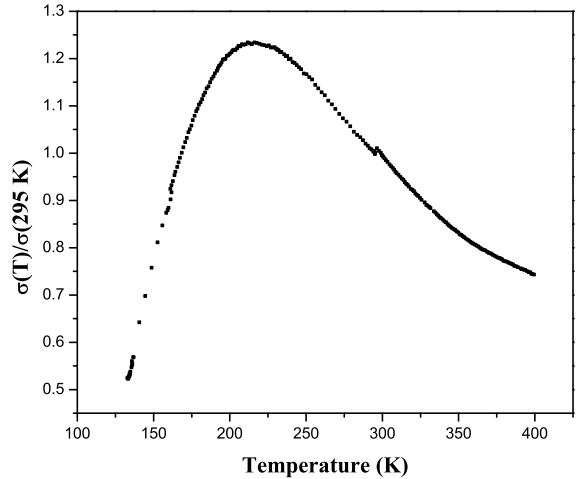}}
\caption{Microwave conductivity at 10 GHz of Mg$_5$C$_{60}$.}
\label{fig:MW}
\end{figure}

Both the microwave conductivity and ESR measurements show that Mg$_{5}$C$%
_{60}$ is metallic above room temperature and becomes insulating at low
temperature. Fig. \ref{fig:MW} displays the temperature dependence of the
microwave conductivity of Mg$_{5}$C$_{60}$ normalized to 295~K. At high
temperatures the conductivity decreases with increasing temperature as it is
usual in metals where scattering is due to phonons. The low temperature
behavior is different from usual metals. The conductivity has a broad maximum and
decreases below 200~K. This behavior corresponds to a smooth
transition between metallic and insulating states.

The ESR spectrum consists of a single Lorentzian line at all temperatures
both at 9 and 225~GHz confirming the phase purity of the sample. Phase
inhomogeneities usually split the line or at least 
inhomogeneously broaden it due to differences in the $g$-factor. If the Mg
concentration were inhomogeneous then the high frequency ESR lines would be
distorted, especially in the low temperature insulating state where spatial
variations are not averaged by spin diffusion.

\begin{figure}[tbp]
\centerline{\includegraphics[width=1\hsize,trim=0 0 12pt 0]{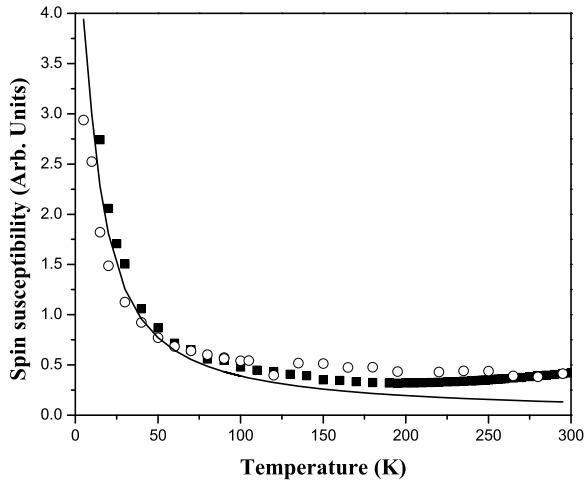}}
\caption{ Temperature dependence of the spin susceptibility of Mg$_5$C$_{60}$. 225 GHz ESR measurements (empty circles) are normalized to the 9 GHz ones (squares) at 300 K. The continuous line is the Brillouin function of
spin $1/2$ in 8.1 T field normalized at 50 K to the 225 GHz measurement.}
\label{fig:ESR1}
\end{figure}

The change from the low temperature insulating to the high temperature
metallic state is observed in
the temperature dependence of the spin susceptibility also (Fig.~\ref{fig:ESR1}). For $T\geq 200$~K the
spin susceptibility is almost independent of temperature, as expected for the Pauli susceptibility of a degenerate electron gas. In normal metals the spin
susceptibility $\chi $ is related to the density of states at the Fermi
level $n(E_{F})$: $\chi =1/4 g^{2}\mu _{B}^{2}n(E_{F})$,
where $g$ is the spectroscopic $g$-factor and $\mu _{B}$ is the Bohr
magneton. At lower temperatures a Curie contribution characteristic of a
small concentration of localized spins dominates the ESR spectrum (Fig.~\ref{fig:ESR1}). The change of the temperature dependence of the static
susceptibility from Pauli to Curie like is gradual, there is no well defined
phase transition. We suggest that the smooth transition to an insulating
ground state is due to Anderson localization. At a certain degree of
disorder of the lattice the electronic wave functions with an energy below
the mobility edge, E$_{c}$, become localized. \cite{Anderson58,Mott67} If
the Fermi energy lies below E$_{c}$ the ground state is insulating.
Electrons are delocalized and the behavior is metallic if the temperature is
higher than E$_{c}$/k$_{B}$.

\begin{figure}[tbp]
\centerline{\includegraphics[width=1\hsize,trim=0 0 12pt 0]{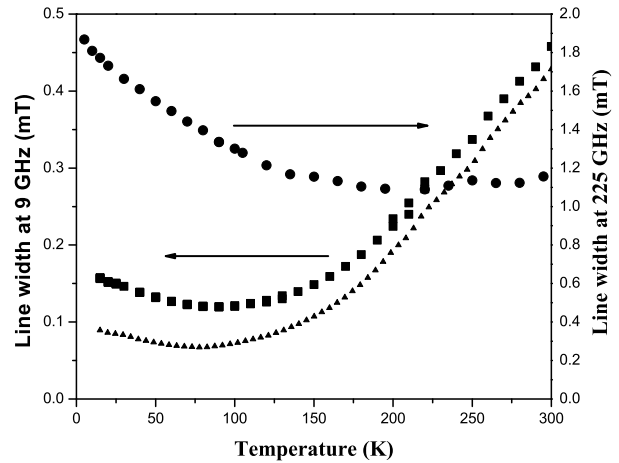}}
\caption{ Temperature dependence of the ESR line width of Mg$_5$C$_{60}$ at 9
GHz (squares) and 225 GHz (circles). The temperature dependence of the
frequency independent component of the line width $a(T)$ is also reported
(triangles)}.
\label{fig:ESR3}
\end{figure}

An additional insight into the electron relaxation mechanism is given by the
frequency and temperature dependence of the ESR line width. The 9~GHz line
is a symmetric Lorentzian with a 4.5~G line width at room temperature. The
line broadens inhomogeneously with increasing frequency. The frequency and
the temperature dependences of the line width may be separated to give $\Delta H=a(T)+b(T)H$. The frequency independent term $a(T)$ is almost
constant below 100~K (see Fig. \ref{fig:ESR3}) and increases linearly with
temperature as in metals in which the main relaxation mechanism is the
electron-phonon interaction.\cite{Yafet63} The term $b(T)$ increases with frequency and with the lowering of temperature. The
distortion of the line and the broadening at 225~GHz is due to a partially
averaged g-factor anisotropy. We attribute the broadening at $T<200$~K (see
Fig.~\ref{fig:ESR3}) to the inhomogeneous interaction between paramagnetic
defects randomly distributed in the lattice.

We measured a lower limit of the depolymerization temperature by searching
for changes in the  ESR spectra after
annealing at higher temperatures. The sealed Mg$_{5}$C$_{60}$ sample was
heated at a rate of 10 K/min to each temperature, kept there for 15 minutes, quenched into water and then the
9 GHz ESR spectra were recorded at ambient temperature. We repeated this cycle twice in 25 K steps between 300 and 848 K. We found that treatment up to 823 K
leaves the ESR spectrum unchanged. At 848 K the ESR spectrum and the
color of the quartz sample holder changed. These changes indicate
a reaction of Mg (diffusing out of the sample) with quartz. It is not clear whether without this reaction depolymerization would take place. We conclude that the polymer phase is stable for at least 30 minutes at 823 K.

\section{Conclusion}

Mg$_{x}$C$_{60}$ with $x=5-5.5$ is a single phase, two dimensional fulleride polymer. X-ray diffraction indicates a finite stoichiometry range with the same structure. Vibrational spectroscopy reveals that the Mg-fullerene interaction is more complicated than simple charge transfer, with a possible covalent component.
Microwave conductivity shows that Mg$_{5}$C$_{60}$ is metallic
for $T$ $>$ 200 K as confirmed by the temperature
dependence of the spin susceptibility and this is the reason for the temperature
dependence of the ESR line width in the metallic state. We attribute the transition from metallic
to paramagnetic insulating states below 200~K to a disorder driven Anderson
localization. The stability of the polymeric structure of Mg$_{5}$C$_{60}$ up to temperatures above 800 K is exceptional, all other polymers of charged fullerides connected by single or double C-C bonds
decompose at much lower temperature.

\section{Acknowledgment}

We gratefully acknowledge G. Oszl\'anyi for helpful discussions and L. Forr\'o for providing the X-band ESR spectrometer at the EPFL . This work
was supported by the Hungarian National Research Fund (OTKA) through grants
No. T 049338, T 046700, TS049881, F61733 and NK60984.

%\bibliography{C60IRb}

\end{document}